\documentclass[%
aip,
jmp,%
amsmath,amssymb,
reprint,%
]{revtex4-1}

\usepackage{graphicx}
\usepackage{dcolumn}
\usepackage{bm}
\usepackage{amsmath}
\usepackage{natbib}

\begin{document}


\title[An \textit{In Silico} Model to Simulate the Evolution of Biological
Aging]{An \textit{In Silico} Model to Simulate the Evolution of Biological
Aging}

\author{A. \v{S}ajina}
\altaffiliation{Friedrich-Wilhelm Gymnasium, Severinstrasse 241, Cologne, Germany.}
\author{D.R. Valenzano}%
\altaffiliation{Corresponding author: Dario.Valenzano@age.mpg.de}
\affiliation{Max Planck Institute for Biology of Ageing,  Joseph-Stelzmann Strasse 9b, 50931 Cologne, Germany.}


\begin{abstract}
Biological aging is characterized by an age-dependent increase in the
probability of death and by a decrease in the reproductive
capacity. Individual age-dependent rates of
survival and reproduction have a strong impact on population dynamics, and the
genetic elements determining survival and reproduction are under
different selective forces throughout an organism lifespan. Here we
develop a highly versatile numerical model of genome evolution --- both
asexual and sexual --- for a
population of virtual individuals with overlapping generations, where the genetic elements
affecting survival and reproduction rate at different life stages are
free to evolve due to mutation and selection.
Our model recapitulates
several emerging properties of natural
populations, developing longer reproductive lifespan under stable
conditions and shorter survival and reproduction in unstable
environments. Faster aging results as the consequence of the reduced
strength of purifying selection in more unstable populations, which have 
large
portions of the genome that accumulate detrimental mutations. Unlike sexually
reproducing populations under constant resources, asexually reproducing populations
fail to develop an age-dependent increase in death rates and decrease in
reproduction rates, therefore escaping senescence. Our model provides a powerful \textit{in silico} framework to
simulate how populations and genomes change in the context of
biological aging and opens a novel analytical opportunity to characterize
how real populations evolve their specific aging dynamics.
\end{abstract}

\maketitle

%

\section{\label{sec:level1}Introduction}
\indent Natural populations evolve from the interaction between external
selective forces and the internal capacity to respond to them. 
External forces include predators, parasites and
available resources, and population response to these forces
importantly depends
on rates of mutation and recombination. Standard evolutionary theory of aging predicts that deleterious mutations affecting early or late survival
are differentially selected, with mutations negatively
affecting survival in early life being readily removed by purifying selection,
as opposed to those affecting survival in late life, which instead have
a tendency to accumulate due to weaker selective forces
\citep{Medawar1952, haldanenew1941, hamiltonthe1966,Rose1991, charlesworth2000}. A consequence of this
phenomenon is that young individuals have lower death rates (i.e. higher chances to survive)
compared to old individuals. Furthermore, it is theoretically possible
that selection could favor alleles that
have a detrimental effect on survival in late life stages, as long as they have a
beneficial effect early on in life -- i.e. alleles that contribute to
an aging phenotype
can be fixed in the gene pool if their overall effect on fitness is
positive \citep{Williams1957}.\\\indent Although these models greatly helped to frame the genetics
of biological aging on a solid evolutionary basis, there are some important
aspects that were left out, which are key for our understanding on how
selection shapes individual fitness in biological populations. In
particular, exclusively focusing on how selection shapes
age-specific survival, they did not analyze how genes that
regulate survival and reproduction might
coevolve and how their interaction could affect individual fitness and
population stability under different selective pressures, which are
key aspects to understanding how natural populations evolve.
\\\indent Unlike analytical methods, computer simulations -- or
numerical models -- can
sustain high parameter complexity and are particularly suited to model
the evolution of dynamic systems, such as biological systems
\citep{Penna1995, Dzwinel2005}.\\
\indent To analyze the role of selection in shaping the genetic basis of reproductive
aging and age-dependent changes in death rate, we developed an \textit{in
silico} model that simulates the evolution of biological populations,
where each individual agent is provided with a genome that defines the
probabilities to reproduce and survive. We let a genetically heterogeneous population
evolve under different environmental constraints, including limiting
resources, sexual or asexual reproduction and different
population sizes. We find that, in a
sexual model, the force
of selection decreases with age after sexual maturation. However, we
demonstrate that, in stable environments, selection favors prolonged survival and
reproduction after sexual maturation more than in unstable
conditions. In unstable environments, characterized by continuous
expansions and compression of population size, we observe
the emergence
of increased early mortality. Importantly, we observe that
the hallmarks of biological aging, i.e. the time-dependent decrease in
individual survival and reproduction, are more dramatic in the sexual model and do not
evolve in the asexual model under constant resource conditions. Our model offers the possibility to
address key evolutionary questions regarding the evolution of biological
aging at both individual and population level, and its applications
can give key insights on how genomes contribute to
the time-dependent deterioration of biological functions.
\section{\label{sec:level1}The Model}
\paragraph{The genome.}
\indent Our model simulates the evolution of a population of agents, which
evolves through a sequence of discrete time intervals, called
\textit{stages}. For each agent $i$, the
probability to surviving and reproducing at each stage is defined by a bit-string code that we
name \textit{genome} ($G_i$). 
\begin{figure}[h]
    \centering
    \includegraphics[width=0.45\textwidth]{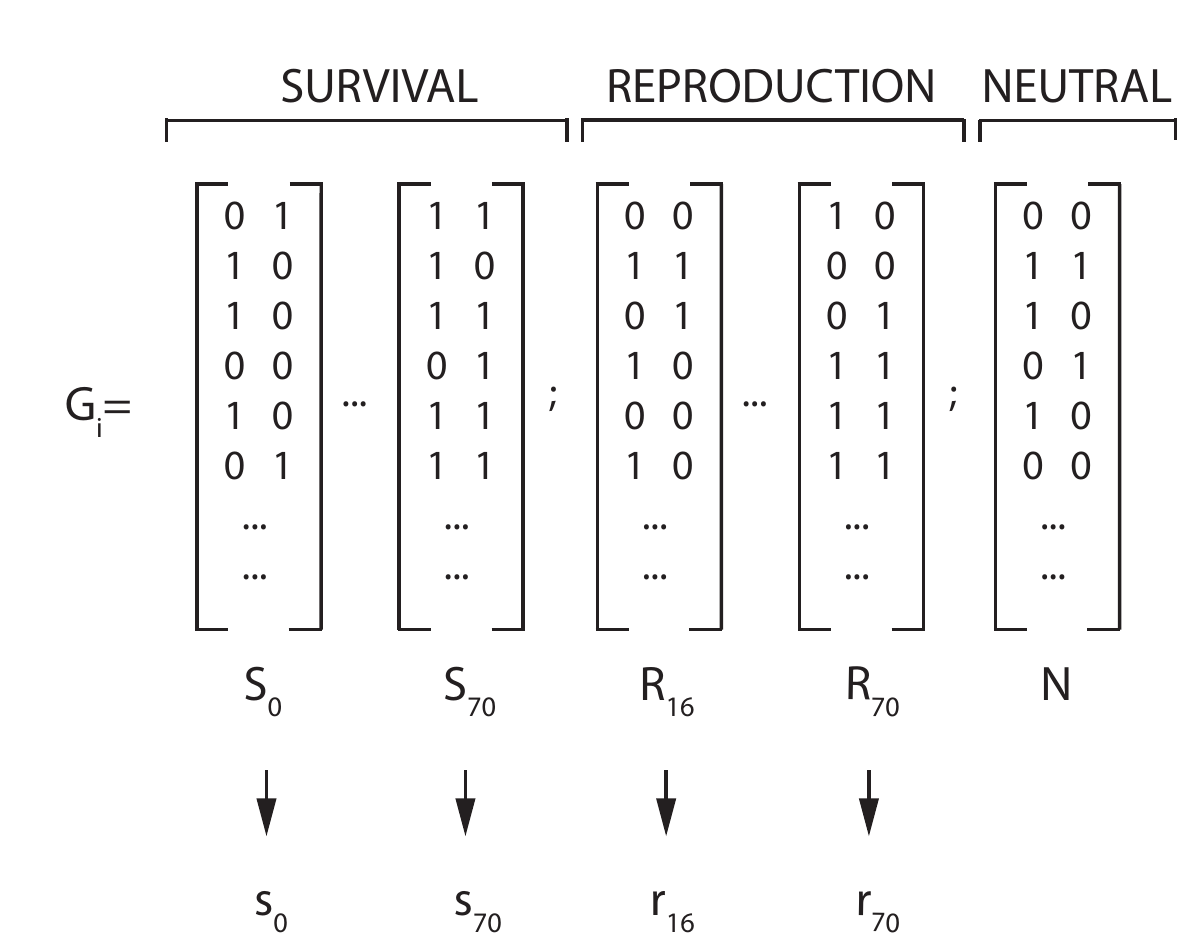}
    \caption{Genome composition. The
      genome ($G_i$) contains a sequence of discrete bit arrays -- $S_i$ and
      $R_i$ -- that define
      the probability to survive from one stage to the
      next and to reproduce at each stage, respectively indicated as
      $s_i$ and $r_i$. Reproduction starts at age 16, therefore there
      are no $R$ and $r$ values with an index smaller than 16. A neutral portion of the
      genome, indicated by $N$, has no corresponding phenotype, and
      is used as a control region to measure neutral evolution.}
\end{figure} 
The probability to surviving from one stage to the next is
indicated by $s_k$, and we refer to it as \textit{transition
  probability}. The transition probability for stage k is
proportional to the numbers of $1's$ present in a corrensponding 20-bits array, called
$S_k$ (FIG. 1). Analogously, the  probability to reproducing at
stage k, indicated by $r_k$, is proportional to the number of $1's$ in the
20-bits array $R_k$:
\begin{equation*}
s_k =
\left(\frac{max_{surv}-min_{surv}}{21}\right)\sum_{b=0}^{20}S_{kb} + min_{surv}
\end{equation*}
\begin{equation*}
r_k =  \left(\frac{max_{repr}-min_{repr}}{21}\right)\sum_{b=0}^{20}R_{kb}+min_{repr}
\end{equation*}
with both $S_{kb}$ and $R_{kb}$ that can be = $0$ or $1$ and $max_{surv}$ and
$max_{repr}$ being the initially set maximum allowed survival and reproduction
probability, respectively.  
\\\indent $G_i$ also contains a bit array that is not
translated in any phenotype (indicated as $N$ in FIG. 1), which is used as a functionally-neutral
portion of the genome.
\\\indent For each agent $i$ in the population, $G_i$ is a continuous
bit array, i.e. all $S_k$, $R_k$ and $N$ are contiguous. For representation purposes, in all the
plots shown in results, both $s_k$ and $r_k$ are represented in
ascending order, i.e. with $s_k$ preceding $s_{k+1}$. However,
the actual order along the bit array is randomized, so that $S_k$ does not
necessarily preceed $S_{k+1}$.
\paragraph{Simulation progression.} 
The seed (or starting) population is composed of 
genetically heterogeneous agents, whose genomes are randomly generated
bit arrays, which have on average an equal number of $1's$ and
$0's$. Each agent from the seed population has a specified
chronological age, and becomes one time-unit older at every new
stage. Resources are used as the population growth-limiting
factor. Resource amount is defined before the simulation starts and, at each stage, the
population can be in a state of resource abundance or shortage, i.e. $N_t \leq R_t$ and $N_t > R_t$, with $R_t$ being resources available at stage $t$ and $N_t$ being the
populations size (i.e. number of individuals) at stage $t$. In the
former situation, for each individual, the death rate will
correspond to $1 - s_t$, while in the latter it will correspond to
$\mu\left(1 - s_t\right)$, with $\mu$ being a constant that increases actual death rate
for all the agents in case of resource shortage. Depending on the
simulation type, resources are set to be either constant or variable. In the
\textit{constant resource} model, resources are the same at every new
stage, regardless from the leftover resources from the previous stage. However, population growth and contraction will affect the
death rate depending on whether population size exceeds resouce
units. In the \textit{variable resource} model, the amount of available
resources is the sum of two quantities. One is proportional to the
difference between resources and population size from the previous
stage (leftover resources), and the
other is a fixed resource increment $\bar{R}$:\\
\newline
$
R_{t+1} = \begin{cases}
(R_t - N_t)k +   \bar{R} & \text{ if } N_t < R_t \\
(R_t - N_t) + \bar{R} & \text{ if } N_t \geq R_t
\end{cases}
$
\\\newline
\indent If the amount of resource units from the previous stage exceeds population size, the leftover
resources regenerate proportionally to a fixed proliferating
value $k$ -- set at the beginning of the simulation -- and are added to the fixed
resource increment $\bar{R}$. However, if the population size exceeds or
is equal to the resource units, i.e. no resource is left from the previous
stage, the
difference between population size and resources available is subtracted from the fixed
resource increment $\bar{R}$. If this difference is $\leq0$,
resources are then set to $0$. In this case, death rate is
affected (see above).
\\\indent Reproduction follows resource consumption (FIG. 2). In the asexual model, each
individual can reproduce or not, based on the probability provided by
the individual age-specific $r_k$ value.
\begin{figure}[h]
    \centering
    \includegraphics[width=0.35\textwidth]{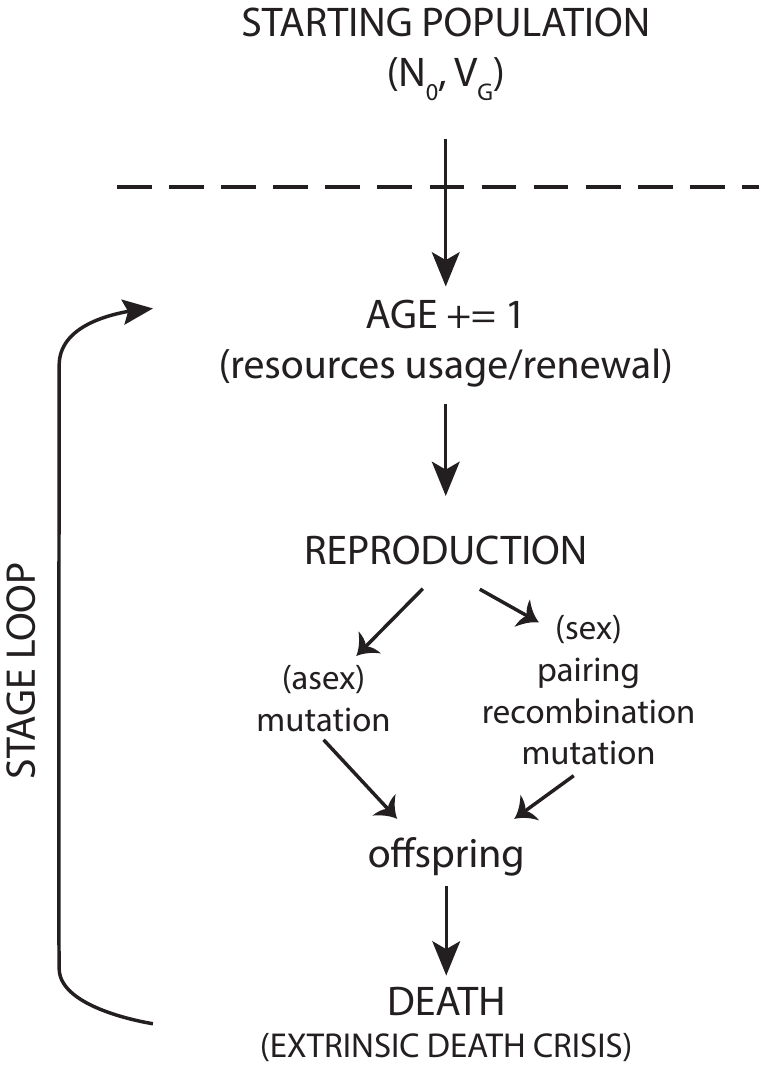}
    \caption{Schematics of events sequence in the simulation. $N_0$ and $V_G$ represent the initial number
      of individuals and the genetic variance initially defined for
      any seed (or starting) population.}
\end{figure} 
In the sexual model, the individuals that, based on
the stage-specific $r_k$ value, are selected to reproduce, undergo
chromosome recombination. Recombination is determined by a fixed
recombination frequency, and is followed by the random selection of one of
the two chromosomes (either the left or the right
in each $S_{0:70}, R_{16:70}$, and $N$ from FIG. 1) to match a
corresponding recombining chromosome selected from another individual.
Chromosome matching is followed by mutation, which is set by a
fixed mutation rate before each
simulation. Mutation transforms $0$s to $1$s and, vice versa, $1$s
to $0$s. Since any $0$ mutating into $1$ in both $R_k$ and $S_k$ translates
into an increase in the probability of reproducing or surviving,
respectively, corresponding to a phenotypically \textit{beneficial} mutation, we let
$0$ to $1$ mutations to be 10 times less likely than $1$ to $0$
mutations, in order to let mutations to be more likely detrimental
than beneficial. 
\\\indent Once the \textit{mutation step} takes place, a novel genome is
complete, and it represents
the new survival and reproduction program for a newborn agent, which
is added to the current population with
the starting age of 0.
\\\indent Following reproduction, each agent's $s_k$ values contribute to
determining whether the agent survives or is eliminated from the
population. The actual 
probability of survival through any stage depends on i) each agent's
$s_k$ value for the actual age $k$ at the given stage, ii) the
difference between resources and population, which can add a weight on
each agent's death risk, and iii) the \textit{extrinsic death
  crisis} parameter, which can be initially set and determines whether
-- at specific stages -- the population undergoes massive death.  
\\\indent All individuals reach sexual maturation at stage 16,
therefore $R_k$ start at $k = 16$.
\subsection{\label{sec:Level2}Fitness}
We define individual \textit{genomic fitness} ($GF_{ind}$) as the
cumulative individual lifetime expected offspring contribution,
given the genetically determined probabilities to survive and
reproduce. For instance, the \textit{genomic fitness} for an agent at age 50
corresponds to the product of two terms, one that defines the
probability to surviving to reproductive age:
\begin{equation*}
S_{RA} = s_0s_1s_2...s_{15}
\end{equation*}
and one that defines the probability to surviving and reproducing at
all the stages that follow sexual maturation, up to the last stage
(age 50 in this example):
\begin{equation*}
SR_{16-50} = s_{16}r_{16} + s_{16}s_{17}r_{17} + ... + s_{16}s_{17}...s_{50}r_{50}
\end{equation*}
with $s$ and $r$ as the transition probability for
survival and reproduction, respectively, defined in the individual
genome by their corresponding $S$ and $R$ values.
\begin{figure*}
    \centering
    \includegraphics[width=0.95\textwidth]{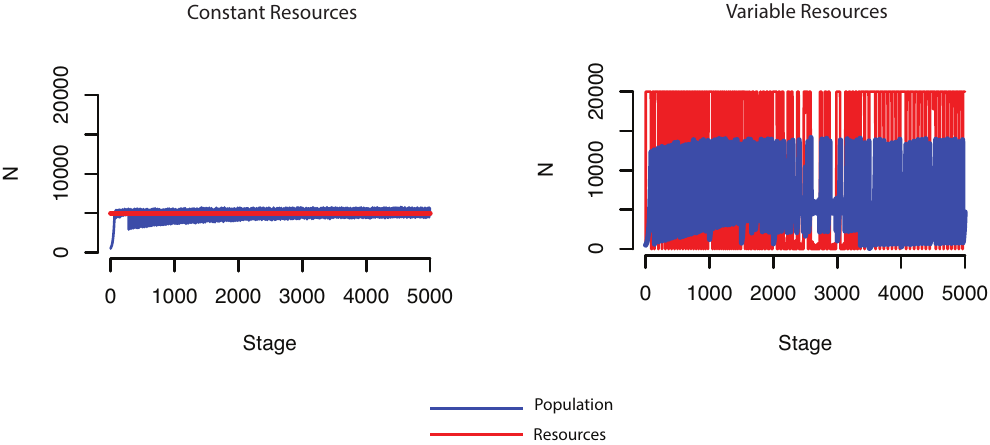}
    \caption{Sample population growth and population size oscillations under
      constant (left) and variable (right) resource conditions. Only the first 5k stages of the
      simulations are displayed for simplicity. In the \textit{constant resource} conditions
      resources are fixed and do not change as a function of
      population size, whereas, in the \textit{variable resource} conditions,
      resources vary depending on an intrinsic growth rate, a fixed
      stagewise increment, and population size.
  }
\end{figure*} 
More generally, this product can be written as:
\begin{equation*}
GF_{ind} = \left(\prod_{k=0}^{15}s_k\right)\left(\sum_{i=16}^{Age}r_i\prod_{j=16}^{i}s_j\right)
\end{equation*}
with $Age$ as the individual age at which fitness is measured.
\\\indent We also define, for any stage, an average population genome
and its associated \textit{genomic fitness} ($GF_{avg}$), as:
\begin{equation*}
GF_{avg} = \left(\prod_{k=0}^{15}\bar{s}_k\right)\left(\sum_{i=16}^{Age}\bar{r}_i\prod_{j=16}^{i}\bar{s}_j\right)
\end{equation*}
with $\bar{r}$ and $\bar{s}$ as the average reproduction and survival
probabilities for each age.
\\\indent Analogously, we can define \textit{relative individual genomic
  fitness} as the ratio of individual fitness to the sum of all the
individual fitness:
\begin{equation*}
RGF_{i} = \frac{GF_{i}}{\sum_{ind=1}^{N}GF_{ind}}
\end{equation*}
\\\indent Tracking population size throughout the
simulation progression, enables to measure rates of \textit{actual survival} at
any stage, for each age class, defined as:
\begin{equation*}
\sigma\textsuperscript{a}_k = \frac{N\textsuperscript{a+1}_{k+1}}{N\textsuperscript{a}_{k}}
\end{equation*}
with $N\textsuperscript{a}_{k}$ as the number of individuals of age $k$
at stage $a$. Replacing $\bar{s}$ with
$\sigma$, allows a corrected $GF_{avg}$,
which we name \textit{population fitness} or $F_p$:
\begin{equation*}
F_p = \left(\prod_{k=0}^{15}\sigma_k\right)\left(\sum_{i=16}^{Age}\bar{r}_i\prod_{j=16}^{i}\sigma_j\right)
\end{equation*}
\section{\label{sec:Level1}Results}
\indent During each simulation, an initial random gene pool is
generated, and $S_{ik}$, $R_{ik}$ and $N_i$ (see FIG. 1) evolve freely due to
mutation and -- in the sexual model --
recombination. However, this model also allows populations to be \textit{re-seeded} from end-stages
of previous simulations. Population size oscillates as a function of
i) available resources (FIG. 3) 
and ii) evolving distributions of age-dependent
reproduction and survival probabilities encoded in each individual
genome, which affect the ratios of births over deaths. The model does
not bias the direction in which populations will evolve higher or lower
survival and reproduction probabilities throughout age, and all the
observed changes in age-specific reproduction and survival probabilities ($r_k$ 
and $s_k$) are due to selection and drift.
\begin{figure*}
    \centering
    \includegraphics[width=0.95\textwidth]{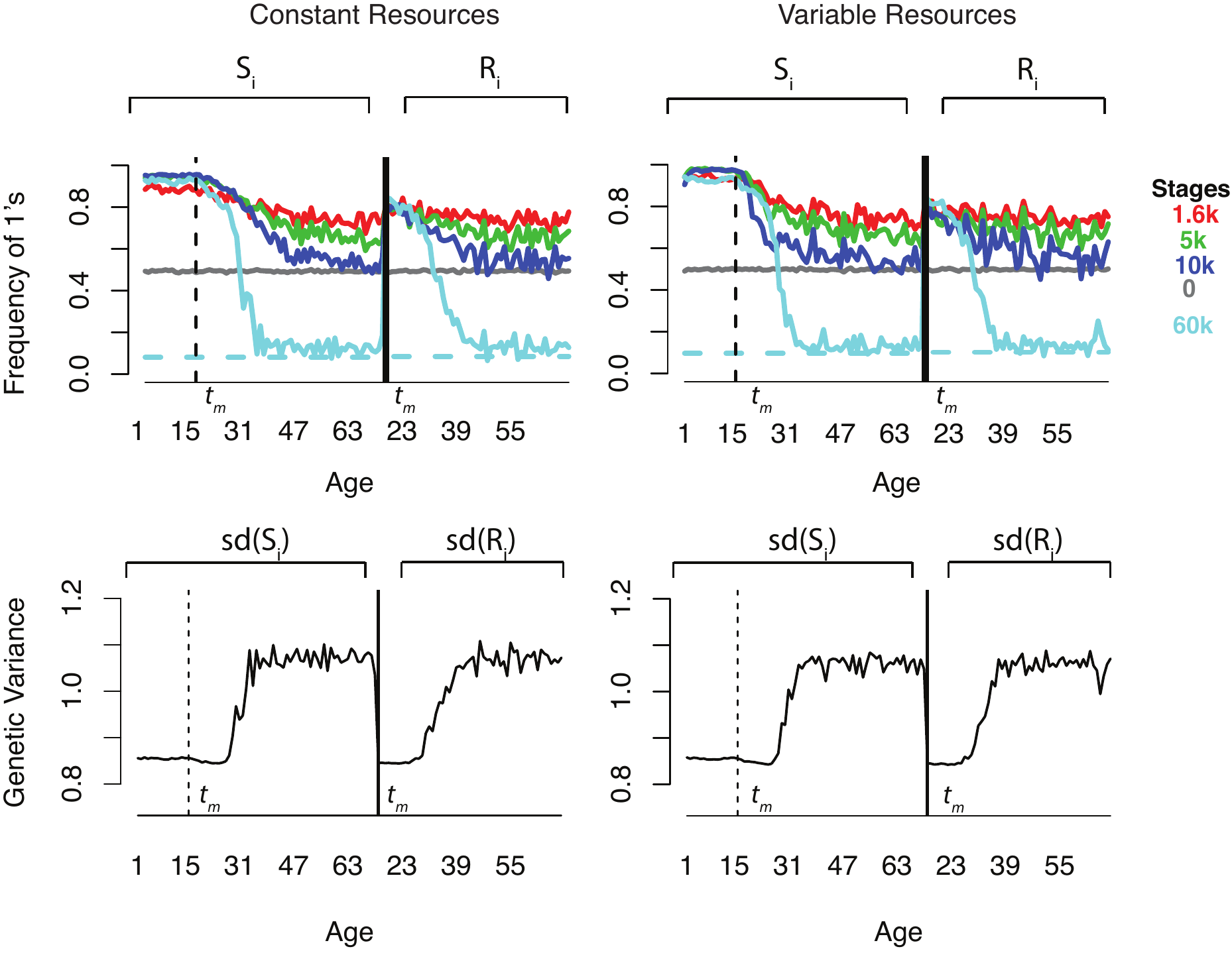}
    \caption{Evolution of the genome in the sexual model. Above: The x-axis
      represents the position of $S_i$ (on the left) and $R_i$ (on the
      right) along the genome, sorted
      according to the age they correspond to. The y-axis represents
      the ratio of 1's for each transition
      probability $S_i$ and $R_i$ indicated on the x-axis. The genomes
     are randomly generated at stage 0 (grey line) and start with an
      even distribution of 0's and 1's, for all the $S_i$ and $R_i$
      (average values only are shown). As the simulation proceeds
      (different color lines), mutation and selection change the
      frequencies of 1's and 0's for each age-specific $S_i$ and $R_i$. The genetic modules relative to
      survival ($S_i$) fix higher frequencies of 1's before and about
      sexual maturation (vertical dashed line and $t_m$), while reproduction
      ($R_i$, not defined before $t_m$) fix higher frequencies of 1's starting at $t_m$. At later stages of the
    simulation, genetic modules for both survival and reproduction
    after sexual maturation accumulate more deleterious mutations than
    at earlier stages, as indicated by the lower values to the right
    of the cyan curve (60k stages). Below: population average of the genetic variance
    for $S_i$ and $R_i$ as a function of age after 60k stages of
    simulation. Genetic Variance before $t_m$ is low, and increases
    for both survival and reproduction with Age $> t_m$. Fixed resources are set to 5k
    units, mutation rate is set to 0.001 per
    site and recombination rate is set to 0.01.
  }
\end{figure*} 
\subsection{\label{sec:level2}Sexual model}
\indent Under constant resources, populations of agents that 
reproduce sexually evolve to a state of higher $S_i$ before sexual maturation
and continuously declining $S_i$ and $R_i$ values when $i >
t_m$, with $t_m$ as the time of sexual maturation (FIG. 4). 
\\\indent Since $s_i$ and $r_i$ are directly proportional to $S_i$ and
$R_i$, respectively, also the age-dependent probabilities of surviving
and reproducing become accordingly higher in early life and decline after sexual
maturation. The phenotypically neutral portion of the genome (indicated by $N$ in
FIG. 1, and by the dashed lines in FIG. 4 and in FIG. 5B), used as a control region
to test the effects of mutation without selection on the genome sequence
evolution throughout the simulation, enables to show that the
distribution of the $S_i$ and $R_i$ values throughout age is under
strong selection before and after sexual maturation, for all
the ages where $S_i$ and $R_i$ values are above the curve
corresponding to the expected values
(horizontal dashed line in FIG. 4 and in FIG. 5B). For each age, the difference
between $S_i$, $R_i$ and the expected values corresponding to the
neutrally evolving portion of the genome ($N_i$), is
proportional to the strength of selection.
\\\indent Population genetic variance for $S_i$ reaches low
values from stage 0 until about sexual maturation, and has its minimum
around sexual maturation also for $R_i$, as a consequence of strong purifying
selection. After the onset of sexual maturation, genetic variance for
both $S_i$ and $R_i$ increases rapidly, reflecting weakened purifying
selection at later ages (FIG. 4).
\\\indent We tested the effects of resource availability on population
size oscillation and genome evolution. Even with constant resources,
population size fluctuates due to the increased overall mortality induced by
population size exceeding available resources (see II.b). In the \textit{constant resource
conditions}, population oscillations amplitudes were comparable when
resources were set to 1k or 5k units (FIG. 5A). However, under \textit{variable
resource} conditions, population size is a function of the
stage-dependent resources increment ($\bar{R}$) parameter. Population
size oscillations had elevated amplitude in
early stages, and stabilized to smaller amplitudes when $\bar{R}$ was high (5k units) (FIG. 5A). However, when $\bar{R}$ was set to 1k units (or
lower), constraining population size to lowever values, population size oscillations did not stabilize,
indicating that smaller populations are more unstable than larger
populations. Importantly, in the \textit{variable resource} condition, if
$\bar{R}$ is large, resources do not oscillate with large amplitude
once the population stabilizes (FIG. 5A). Decreased oscillation
amplitudes in both population size and resource units is due to $S_i$ and
$R_i$ values evolving towards a stable equilibrium (FIG. 5A).
\begin{figure*}
    \centering
    \includegraphics[width=0.95\textwidth]{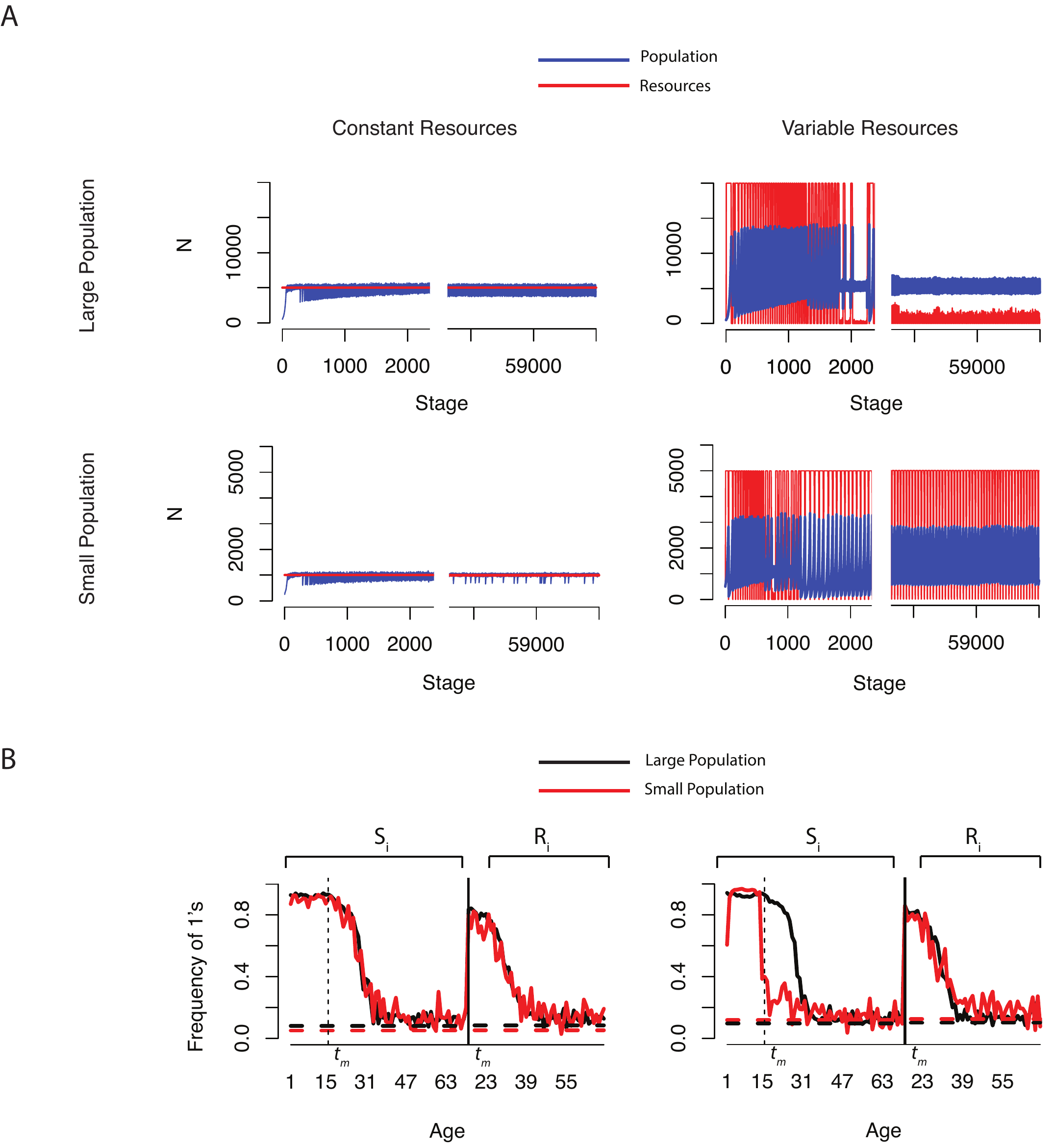}
    \caption{Population, resources and genomic dynamics at different population
      sizes in the sexual model. A: Population and resource dynamics
      in larger and
      smaller populations, under constant (left) or variable
      (right) resource conditions. Population and resource values are
      displayed on the y axis. Resources are shown in red and
      population in blue. Each plot is split in two parts, with early
      stages on the left, and late stages on the right. Each run
      consists of 60,000 stages. Large and small populations are determined by
      the resource value, set to 5k and 1k resource units,
      respectively. In the \textit{constant resource} condition, the
      resource value is fixed, whereas in the \textit{variable resource}
      condition the population size is directly affected by the fixed resource
      increment $\bar{R}$. For sake of representation, $\bar{R}$ is
      not added to the resource values
      on the plot. B: Average $S_i$ and $R_i$ values at the
      last stage of the simulation (stage 60,000) in the constant
      (left) and variable (right) resource condition model. The
      horizontal dashed lines represent the expected $S_i$ and $R_i$ values for for the portion of the
      genome that is not subject to selection, used as a control
      region, indicated by $N_i$ in our model. Mutation rate is set to 0.001 per
    site and recombination rate is set to 0.01. $t_m$ corresponds to the age of sexual maturity. 
  }
\end{figure*} 
This was not observed when resources were variable and $\bar{R}$ was
small, in which case also the age-dependent distribution of $S_i$ in
the post-reproduction phase was lower than in larger populations or in
the \textit{constant resource} condition (FIG. 5B), highlighting that
in our model
genome evolutionary dynamics not only depend on resource availability
(constant or variable), but also on population size. Interestingly, we observed
that smaller populations, in the \textit{variable resource} condition,
evolved to a status of increased early mortality, even
before sexual maturation. This result was not a rare event and emerged in
all our simulations. Additionally, in our model the age-dependent
reproduction evolves in a different fashion from
survival as -- in the \textit{variable resource} condition --
$R_i$ have the same behavior in small and large populations (FIG. 5B,
right). In fact, while $S_i$ drop right after $t_m$
in small populations under variable resources, $R_i$ dynamics are
superimposable in large and small populations (FIG. 5B). 
This behavior might be explained by the fact that in our model only survival is penalised by a
factor $\mu$ when
population size becomes larger than available resources (see II.b).
\subsection{\label{sec:level2}Asexual model}
In the asexual model, we asked what the effects are of population size
and availability of constant or variable resources on population dynamics
and genome evolution (FIG. 6), and how these dynamics are different
from the sexual model. In
the \textit{constant resource} condition, population oscillations decreased amplitude during long simulations (60k stages) both in larger and in smaller
populations (FIG. 6A). Interestingly, unlike what we observed in the sexual model, in the
\textit{variable resource} condition, small-scale population size differences
did not result in balanced equilibrium between resources and population
size in the \textit{variable resource} condition (FIG. 6A). However,
we did observe an increase in the oscillation period between early and
late stages of the simulation, as later stages of the simulation have
longer oscillation periods than earlier stages. Additionally, the
genome of asexually reproducing populations, both in the constant and,
to a lesser degree,
in the \textit{variable resource} condition, did not reach elevated values of
$S_i$ and $R_i$ both in large and small populations -- at least in the
tested population-size ranges -- suggesting
that, in populations reproducing asexually, selection for increased early
survival and reproduction was weaker or had a slower evolution than in populations
reproducing sexually (FIG. 6B).
\subsection{\label{sec:level2}Survival and death rates}
We calculated the observed death rate in both the
sexual and asexual model, for the constant and
variable resource model, in small and large populations (FIG. 7). 
\\\indent Since in our model resources affect survival probabilities
when the population size exceeds the available resources (see II.b), actual
(i.e. observed) mortality at a specific time point is expected to differ from the
mortality predicted solely by the distribution of the $s_k$ values. 
\\\indent In the
sexual model, both in the \textit{constant resource} condition and in the
\textit{variable resource} condition, once reached equilibrium
between population and resources, the age-dependent death rate remained
low before sexual maturation and increased after sexual maturation in
large populations with slightly different dynamics, as in the \textit{constant resource}
condition the increase in mortality happened later than in the
\textit{variable resource} condition (FIG. 7). Small populations in the
\textit{variable resource} condition evolved high early mortality --
i.e. before sexual maturation -- but had an overall slower increase in
the age-dependent death rate, as shown by the smaller slope in the
death rate (FIG. 7). 
\begin{figure*}
    \centering
    \includegraphics[width=0.95\textwidth]{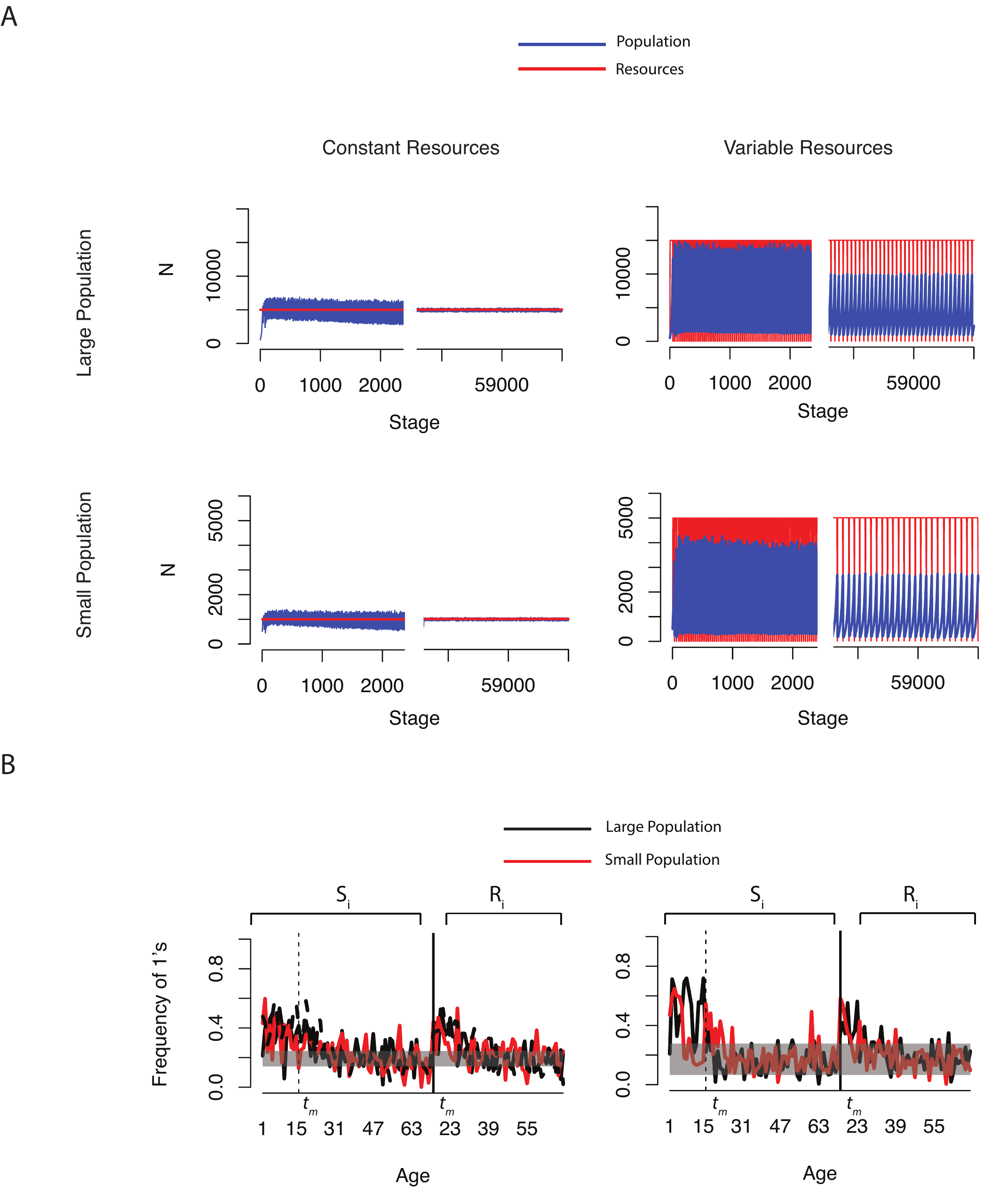}
    \caption{Population, resources and genomic dynamics at different population
      sizes in the asexual model. A: Population and resource dynamics
      in larger and
      smaller populations, under constant (left) or variable
      (right) resource conditions. Population and resource values are
      displayed on the y axis. Resources are shown in red and
      population in blue. Each plot is split in two parts, with early
      stages on the left, and late stages on the right. Each run
      consists of 60,000 stages. Large and small populations are determined by
      the resource value, set to 5k and 1k resource units,
      respectively. In the \textit{constant resource} condition, the
      resource value is fixed, whereas in the \textit{variable resource}
      condition the population size is defined by the fixed resource
      increment $\bar{R}$. B: Average $S_i$ and $R_i$ values at the
      last stage of the simulation (stage 60,000) values in the constant
      (left) and variable (right) resource condition model for small
      and large populations. In the \textit{constant resource} model we added a
      third plot (dashed black line) for \textit{very large} populations, with 25k
      units of fixed resource. The horizontal shaded area represents the interval
      of values for $S_i$ and $R_i$ for the portion of the
      genome that is not subject to selection, used as a control
      region. Mutation rate is set to 0.001 per
    site and recombination rate is set to 0.01. $t_m$ corresponds to the age of sexual maturity. 
  }
\end{figure*} 

\indent In the asexual model, under constant resources, we find no
age-dependent change in the rate of death both in large and small
populations, except for an increase in death rate after age 50 in
small populations (FIG. 7). This might result from the evolution of $S_i$ in the asexual
model under constant resources, which does not significantly
deviate from the expected values reached in the control genome region
(FIG. 6B, left). However, the constant death rate across all the
ages in the \textit{constant resource} condition is associated with
extemely high
population oscillation stability, as indicated by the limited
population size oscillation amplitude (FIG. 6A). Interestingly, we do observe an age-dependent increase in the
rate of death in both the small and large populations for the \textit{variable
resource} model (FIG 7). This behaviour is associated with high amplitude in population
size oscillation, which in our model leads to increased overall
mortality (FIG. 6A).
\begin{figure*}
    \centering
    \includegraphics[width=0.95\textwidth]{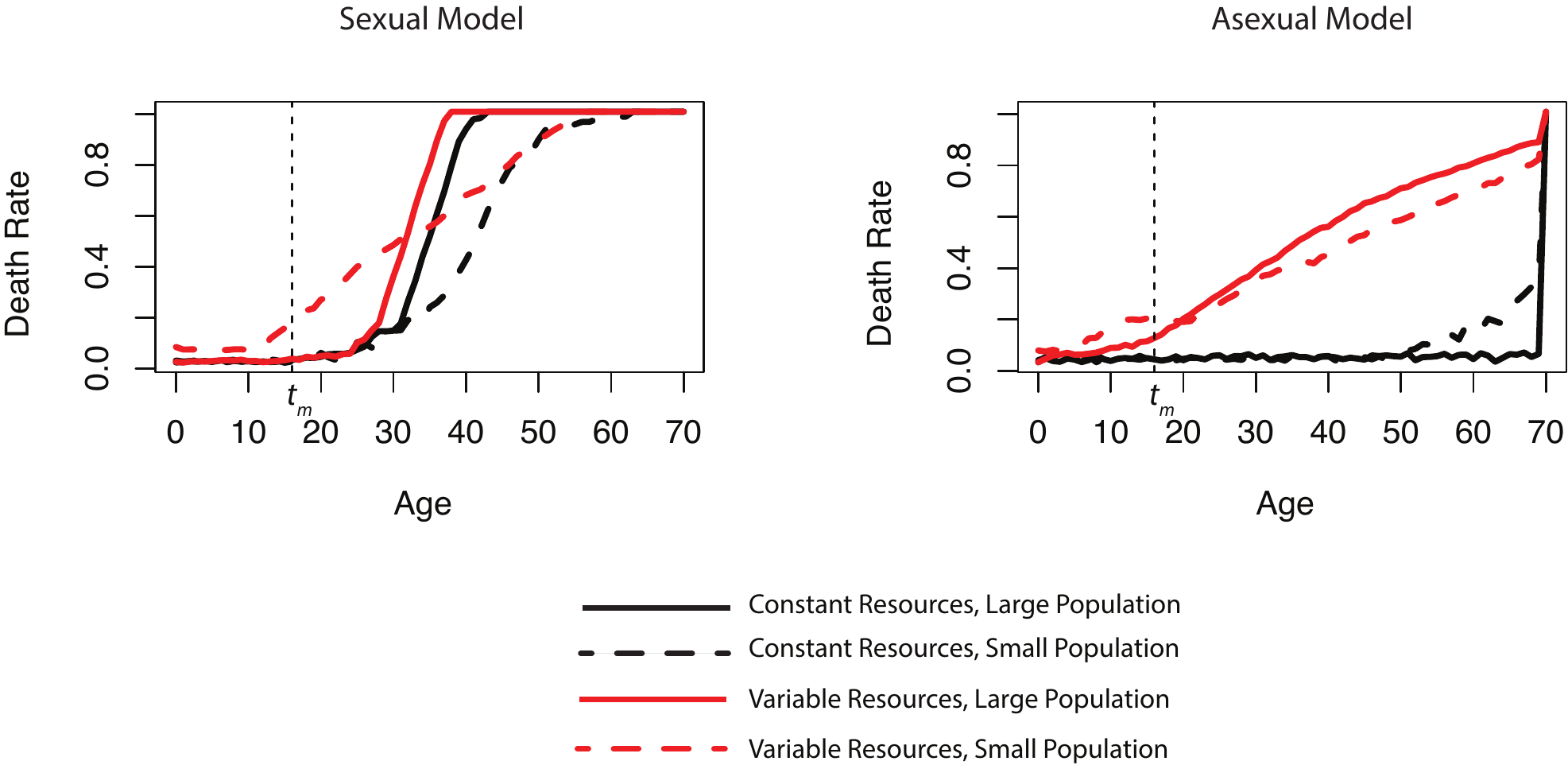}
    \caption{Observed death rate in the sexual and asexual models, with
      constant and variable resources, in large and small
      populations. Death rate is expressed, for each age $a$ at stage $s$, as
      $\frac{N(a)_{s}-N(a+1)_{s+1}}{N(a)_{s}}$, with $N(a)_s$ as the
      number of individuals of age $a$ at stage $s$. The displayed values
      are the average death-rates for the last 100 stages of 60k
      stages-long simulations. $t_m$ corresponds to the age of sexual maturity. 
  }
\end{figure*} 

\section{\label{sec:level1}Discussion}
Studying the changes in individual probability of survival and reproduction is the
focus of gerontology, experimental aging research and evolutionary
biology. Aging research is ultimately
focused on identifying the molecular mechanisms that affect the changes in the probabilities of surviving and
reproducing throughout one individual's lifetime. On the other hand, evolutionary
biology -- and in particular population genetics -- is interested
in the genetic changes that affect the
probabilities of survival and reproduction over several generations, as they directly determine biological
fitness, i.e. how likely it is to transmit 
specific alleles from one generation to the next.
\\\indent In our model, we combine the ``biology of aging'' and the ``evolutionary biology''
perspectives on survival and reproduction and develop a novel \textit{in
silico} system
where distinct genomic elements separately determine the probabilities
of surviving or reproducing at a specific life stage. Using this approach, we
can investigate how natural selection molds the allele frequencies that
determine survival and 
reproduction at different life stages within the lifetime of an
individual -- as well as throughout generations -- in the
population's gene pool.
\\\indent We let populations evolve either sexually or asexually,
under the effect of mutation and recombination (in the sexually reproducing populations). As populations
evolve under fixed or variable resource conditions, we monitor population and genomic dynamics throughout the simulation. In particular, we monitor the 
evolution of differential survival and reproduction probabilities
at the different individual ages allowed by our
model. We compare the age-dependent values that survival and
reproduction probabilities reach throughout the simulation with those
achieved in a neutral portion of the genome that is not subject to
selection. This enables us to quantify the strength of selection, and
to identify the portions of the genome that are under stronger selection,
i.e. those that
deviate from the neutrally evolving parts of the genome. We find that
populations reproducing sexually evolve gene-pools that 
have high survival probabilities before sexual maturation and progressively lower
survival probabilities after sexual maturation. This is in line with
the expectation that early-acting deleterious mutations are
efficiently removed by
selection from the gene pool \citep{Medawar1952}. Importantly, genomic variance for genetic
elements affecting survival is lower in the pre sexual-maturation
phase -- indicated by $t_m$ throughout the paper -- indicating strong
purifying selection in early life stages. Additionally, the genetic elements affecting early
survival accumulate high levels of beneficial
mutations. Importantly, in our simulation the probabilities to survive 
evolve low values after sexual maturation and the genetic variance for survival (as
well as for reproduction), increases dramatically after sexual
maturation, strongly indicating that the age-dependent
decreased probability of surviving and reproducing, i.e. aging, is
directly correlated with increased genetic variance, which results from
the relaxed purifying selection to maintain elevated
survival and reproduction in late-life stages. This result suggests that genes or gene
variants affecting
late-life survival and reproduction are expected to be associated with
higher population variance compared to genes affecting early-life phenotyes, such
as development and early reproduction.\\\indent In the sexual
model, populations evolving under variable resources
undergo large oscillations in population size during the first few
thousand simulation stages. However, at later stages, oscillations in
population size and resources decrease in amplitude and reach an
equilibrium (FIG. 5A). Additionally, we provided environmental
perturbations to populations ``in equilibrium'' by stochastically
imposing 90\%\ mortality to the whole population. This had no effect on
population size oscillation (data not shown), i.e. the
population-resource equilibrium re-established rapidly after the perturbation. Once reached
this stable equilibrium between resources and population size oscillations,
the \textit{variable resource} model behaved
like the \textit{constant-resource} model. In future studies we will
investigate how varying the \textit{fixed resource increment}
$\bar{R}$ affects population
oscillations and genome composition. Intriguingly, we
observed
that population size affects the probability to reach a
resource-population equilibrium in the \textit{variable resource} model,
as smaller populations do not reach such balance. Larger populations display therefore
higher stability and genetic plasticity compared to smaller
populations. Interestingly, we find that sexually reproducing
populations under \textit{variable resource} conditions
develop increased early-life \textit{infant} mortality, i.e. the survival probabilities in the ages preceding
sexual maturation significantly drop under two conditions: i) when population size is
small and ii) in early stages of large population simulations, before
the population-resource equilibrium is reached
(FIG. 5B and FIG 7). Elevated initial \textit{infant} mortality is associated with population instability
and disappears when population and resource oscillations eventually
balance each other. We find the increase in early mortality paradoxical, since the
early ages preceeding sexual maturation, which in our model starts at age 16,
should be under equally strong purifying selection, and any
deleterious mutation affecting survival probabilities in this life
stage should be rapidly removed from the gene pool. At the same time,
we are aware that many species -- including our own -- display
elevated infant death rates, which then decline to a minimum
around puberty. After puberty, death rate starts raising, as already observed
by R.A. Fisher \citep{Fisher1930}. A possible interpretation for this
surprising elevated \textit{infant mortality} is
that, in the \textit{variable resource} condition, large population
oscillations, characterized by continuous population contractions and
expansions, weakens the action of selection to efficiently
remove deleterious mutations acting at early ages. Additionally,
deleterious mutations become more likely to reach high frequency in
the population due to continuous bottlenecks \citep{Hughes2009}. Alternatively,
it could be speculated that a potentially active mechanism could favour the fixation of early-acting deleterious
mutations to buffer large population oscillations by lowering
the population growth rate, ultimately benefiting population
stability. However, these hypotheses need
to be rigorously tested in future studies.
\\\indent Unlike populations reproducing sexually, asexually
reproducing populations under \textit{constant resource} conditions do not
evolve in our model an age-dependent differentiation in the genomically-encoded 
probabilities to survive and reproduce. In particular, we did not
observe higher probability of surviving or reproducing in early
life stages, compared to late life stages. This
phenomenon can be observed at the genomic level, where the genomically-encoded
survival and
reproduction probabilities do not dramatically differ before and after
sexual maturation (FIG. 6B). Similarly to what seen in the $S_i$, the measured death-rates show -- particularly in the \textit{constant
  resource} condition -- no variation (i.e. do not
increase) with age (FIG. 7). Since we define 
aging as an age-dependent increase in the probability of death
and as a reduction in the probability of reproduction, asexually reproducing
populations -- particularly under \textit{constant resource}
conditions -- seem to not undergo aging. In our model, asexually
reproducing populations under fixed resource conditions are associated to the absence of
senescence. This phenomenon, which spontaneously emerged in our simulation, is surprisingly in line
with the evidences supporting extreme longevity and the lack of
senescence in several organisms
that reproduce clonally \citep{Garcia-Cisneros2015}. However, some
asexually reproducing species are also reported to undergo senescence
\citep{Martinez1992}. Interestingly, our model shows to some degree a detectable
increase in the age-dependent death-rate in asexually reproducing populations under the
\textit{variable resources} condition (FIG. 7). In fact, 
both in large and small populations -- using the same resource and
population ranges
used for the sexually
reproducing populations -- population size and resource units oscillations
did not evolve to a status of equilibrium characterized by lower
oscillation amplitude. This result can be interpreted
as a lower stability of populations reproducing asexually, compared to populations
reproducing sexually, under varying resource
conditions. In future studies it will be important to explore how sexually and
asexually reproducing populations compete against each other under
different resource conditions. 
\subsection{Conclusions} 
For aging to be
considered an adaptation, we
would expect a decreased genetic variance for the genetic
modules that determine late-life decrease in survival and reproduction.
In contrast with this expectation, we observed an age-dependent increased genetic
variance following sexual maturation, which can be interpreted as decreased
purifying selection in late-life phenotypes.
The fitness effect of beneficial mutations acting in late-life is negligible compared to mutations affecting survival and
reproduction at earlier ages. Even in our model, which does not constrain survival and reproduction
to become higher or lower at different life-stages, individual
fitness was not maximized, as survival and reproduction probabilities
were elevated only in early life, before and around sexual maturation,
and then decayed after sexual maturation, due to the effects of relaxed
purifying selection and consequent mutation load on late-life acting
genetic modules. This finding is in line with previous studies
that predicted a declining force of natural
selection throughout lifespan \citep{Medawar1952, haldanenew1941, charlesworth2000}. Since the genetic elements acting on
survival and reproduction in late life-stages accumulated more
deleterious mutations, compared to those responsible for survival
and reproduction in earlier
life-stages,
we observed that a large portion of the genome that affected individual 
survival and reproduction at later ages was strongly associated with
high genetic variance. 
\\\indent In our model, the genetic modules that determine
the probabilities to survive and reproduce at different time
stages are free to evolve due to mutation,
recombination (in the sexual model) and
selection. The simulations analyzed in this study were
run at fixed recombination and
mutation rates (indicated in the figure captions). Future studies will clarify how recombination and mutation rate optima
evolve under different conditions and affect the age-dependent
probabilities of survival and reproduction.
\\\indent To note, our model does not allow for pleiotropic effects
of genetic modules acting at different
ages, since each genetic module is age-specific. We can hypothesize
that in the presence of antagonistic pleiotropism, increased late-life mortality and decreased late-life
reproductive potential could be potentially positively selected, granted that they
provide a fitness advantage in early life stages.
\\\indent Importantly, our model does not allow epistasis to take
place in a direct genetic way, i.e. in our simulations there are no gene
to gene interactions that lead to non-additive phenotypic effects, as each genetic
module is independently responsible for its own phenotypic
output. Although epistasis plays a key role in real
biology, we deliberately did not implement it in our
model for computational simplicity. The genetic
modules for survival and reproduction that are implemented in our
model, rather than being an \textit{in silico} version of individual genes,
can be understood as approximations of gene networks or pathways that influence
either survival or reproduction at given ages. In this
respect, epistasis might not be 
applicable to our model. Importantly, despite the absence of
epistasis, our model reproduces several features of real
population aging, such as increased early survival and rapid increase
of the death rate and decrease of fertility after sexual maturation. 
Despite the absence of \textit{sensu strictu} epistasis and
pleiotropism, this model allows complex coevolution
patterns between genetic modules that affect survival and reproduction
at different life-stages, and provides an indirect test for
the necessity of genetic epistasis and
pleiotropism for the 
evolution of genetically-encoded biological aging. In other words,
since our model evolves features typical of biological aging
populations without implementing epistasis and pleiotropism, we can
conclude that epistasis and pleiotropism -- although important in
organism aging processes -- might not be necessary for the
development of biological aging. Additionally, adopting a model that associates to individual ages discrete
probabilities to survive and reproduce, enables to accurately
calculate fitness at both individual and population level. Fitness
measures factor in
both survival and reproduction probabilities encoded in the genome, as
well as observed survival and reproduction outputs. The study of how
fitness varies throughout population evolution under different
resource conditions and population size will be
further developed in future studies. 
This model provides on one hand a powerful \textit{in silico} system to generate hypotheses to test
on real-life data, and on the other hand provides a numerical platform
to test parameters
derived from real biology, such as measured mutation and
recombination rates, population size, resource availability and
overall extrinsic risk of mortality. The use of this novel model to
simulate the genome evolution in the context of biological aging
opens new possibilities to understand how real populations evolve
such a wide range of life history trait strategies.  
\section{Authors Contribution and Acknowledgements}
D.R.V. developed the model, contributed to
the code and wrote the manuscript. A.S. wrote the
python code and critically contributed to the improvement of the
model.\\\indent We thank Fabio Iocco and Dmitri Petrov for the insightful
suggestions and all the members of the Valenzano laboratory at the Max
Planck Institute for Biology of Ageing for their constructive comments
on the manuscript.\\\indent This work has been entirely supported by the Max
Planck Society and by the Max Planck Institute for Biology of Ageing
in Cologne, Germany.
\section{Bibliography}
\bibliography{preprint3}

\end{document}